# Optimizing Photometric Light Curve Analysis: Evaluating Scipy's Minimize Function for Eclipse Mapping of Cataclysmic Variables

[1]**Anoop Kumar,** [2]**Madan Mohan Tito Ayyalasomayajula,** [3]**Dheerendra Panwar,** [4]**Yeshwanth Vasa**

[1]Independent Researcher, CA, USA; Email: Anoop.kumar.2612@gmail.com

[2]Aspen University, Phoenix, Arizona, USA; Email: mail2tito@gmail.com

[3]Independent Researcher, CA, USA; Email: dheerendra.panwar@ieee.org

[4]Independent Researcher, Milwaukee, WI, USA; Email: yvasa@ieee.org

**Abstract**

With a particular focus on Scipy's minimize function—the eclipse mapping method is thoroughly researched and implemented utilizing Python and essential libraries. Many optimization techniques are used, including Sequential Least Squares Programming (SLSQP), Nelder-Mead, and Conjugate Gradient (CG). However, for the purpose of examining photometric light curves—these methods seek to solve the maximum entropy equation under a chi-squared constraint. Therefore, these techniques are first evaluated on two-dimensional Gaussian data without a chi-squared restriction, and then they are used to map the accretion disc and uncover the Gaussian structure of the Cataclysmic Variable KIC 201325107. Critical analysis is performed on the code structure to find possible faults and design problems. Additionally, the analysis shows how several factors impacting computing time and image quality are included—including the variance in Gaussian weighting, disc image resolution, number of data points in the light curve, and degree of constraint.

## 1. Introduction

The peculiarities of dwarf novae[1] are best understood in light of mass transfer and the properties of accretion discs[2]. Light from their accretion discs, which results from mass transfer, is what distinguishes dwarf novae. This happens when mass bonded to a denser companion star (usually a white dwarf) gravitationally unbinds it from its original star. A hot, narrow disc with dynamic thermal characteristics that produces large bright outbursts is formed within the orbital plane by the accreting matter from the secondary star surrounding the white dwarf. The first known dwarf nova occurred in 1885, although astronomer [1] did not make substantial advances in our understanding of these systems until 1974. In contrast to classical novae, [1] suggested that the luminosity of the accretion disc was the source of the periodic bright outbursts seen in dwarf novae. The Tidal Disruption Model (TDM), which suggests that these outbursts are driven by instabilities inside the accretion disc, was developed as a result by [2].

Generally, a surge of mass transfer from the secondary star and an increase in brightness result—from the disk's viscosity—which is altered when the gas inside it reaches a critical temperature. However, viscosity is the internal friction that changes the directed orbital motion of the gas into random thermal motion. The accretion process itself involves the conversion of the disk's orbital motion into heat through internal friction, causing the gas to spiral inward towards the white dwarf. This phenomenon is poorly understood due to the inadequacies of the weak particle interaction model in explaining the resultant heat generation. Alternative explanations, such as turbulence from convection and random gas motions—are now considered to provide a more accurate model of these outbursts. However, from a theoretical standpoint, an accretion disk behaves similarly to a black body, emitting a continuous spectrum at each radial distance. This emission profile offers insights into the disk's temperature distribution and overall structure. Despite the disk's relatively minor mass compared to the primary or secondary star, it is significantly influenced by the primary white dwarf. This is possible because the luminosity and temperature relationship across the disc can be expressed using the Stefan-Boltzmann law[3]. This formulation serves as the basis for a comparison of temperature profiles between various accretion disc models, which facilitates the detection of abnormalities in the disc or possible flaws in simulation techniques.

---

[1] Dwarf novae are binary star systems in which accretion causes periodic explosive outbursts from a white dwarf and a companion star.
[2] Accretion discs, which power cosmic phenomena, are whirling rings of gas and dust centered on a black hole or star.
[3] According to the Stefan-Boltzmann law, a black body's total energy output is directly proportional to the temperature to the fourth power.

However, eclipse mapping in this situation serves purposes beyond academic curiosity. By taking advantage of dwarf novae's geometric alignment with respect to Earth, eclipse mapping is a method used to learn more about the behaviors and architecture of accretion discs. This technique uses photometric data, which is easier to get hold of and may be obtained via databases such as Kepler's MAST[4]. It is an alternative to spectral tomography. With the help of such data, scientists can investigate the dynamics of accretion discs in systems like those previously investigated by [3] and [4]. Understanding the system's orbital mechanics—which can be obtained through light curve analysis—is necessary for the use of eclipse mapping. However, to further improve the scalability and speed of this research strategy—big data tools like Apache Spark and cloud computing, together with the flexibility of Python and its libraries, can be combined [5]–[7]. Because of these technological advancements, a wider variety of dwarf novae can be studied more thoroughly and efficiently, allowing for faster testing and parameter optimization [8]–[15]. For this field of study to advance, eclipse mapping presents a number of important questions. Comprehending the unique characteristics and potential of eclipse mapping, assessing the consequences of both ideal and suboptimal parameter configurations, and investigating techniques to improve and optimize the approach through the use of contemporary computational instruments and open-source development frameworks are some of these.

The study is as follows; the background will be seen in the following section. The related works are presented in Section 3. The materials and methods are covered in Section 4. The experimental analysis is carried out in Section 5. The discussion portion is covered in Section 6, and Section 7 offers some conclusions and ideas for future research.

## 2. Background

By using the eclipse geometry of binary star systems—eclipse mapping—a technique made popular by researchers like as [16]—allows one to examine the brightness distribution of accretion discs. The assumptions of this method are that the secondary star's surface follows its Roche Lobe Potential[5] (RLP)—that disc material contributing to brightness stays inside the orbital plane, and that radiation released is constant throughout orbital phases. However, this last assumption may make it more difficult to analyze anomalies like outbursts. Therefore, with this method, the most likely brightness distribution over the disc is estimated by using a maximum entropy equation in conjunction with the creation of an eclipse map that depicts the eclipse's geometry at each stage of orbit. Using a grid assigned to the white dwarf's orbital plane, light curve data is visualized on the eclipse map. However, all visible pixels' contributions are added together and their eclipse status is taken into account to determine the disk's flux at any given phase. In order to establish if a pixel is hidden from Earth's view, the model evaluates each pixel's visibility by taking into account the secondary star's orbital location and radius. To effectively simulate the light curve and comprehend the structure of the disc, this visibility is essential. Therefore, in order to measure the information loss that occurs when one probability distribution approximates another—the maximum entropy model applies the idea of the Kullback–Leibler Divergence (KLD) to the disk's brightness distribution. The most likely distribution of pixel intensities with relation to a default image is determined by maximizing this entropy function. In order to determine the most likely image of the disc, the intensity of each pixel is normalized against the sum of the intensities in the default and real images. Improvements in computing power and the availability of more data are what propel eclipse mapping's continuous improvement. Therefore, subsequent research should be conducted on the assumptions regarding the homogeneity of radiated radiation and the precise implementation of visibility models. Our knowledge of star development and the physical dynamics within accretion discs in cataclysmic variables such as dwarf novae will be improved by improvements in the resolution and accuracy of eclipse maps. All of the model's improvements add to our knowledge of the universe and advance the discipline of astrophysics by offering new insights into the intricate mechanisms controlling mass transport and disc luminosity. These improvements could lead to more accurate forecasts and illustrations of the complex processes found in binary star systems.

## 3. Related Works

The term "planetary mapping" refers to a range of methods used to observe a planet's spin, orbit, or spectrum with the primary goal of revealing the planet's atmosphere and structural properties. Technological developments have led to a considerable evolution in this discipline, which has been further boosted by investigations centered on exoplanets. Orbital phase curve mapping is a popular technique that looks at an exoplanet's hemispheres sequentially as they become visible via the planet's orbit. This method, with major contributions from scholars like [17], has proved crucial in describing aspects like day-night brightness differences. Using a star's occulting edge as a natural mask during a planet's secondary eclipse—marked a significant breakthrough in planetary mapping, enabling comprehensive two-dimensional emission mapping of exoplanets. Initially assessed by [18]—this method was subsequently successfully applied by [19]. These mappings are important because they allow us to compare observed brightness distributions with atmospheric circulation model predictions, which shed light on the atmospheric features of exoplanets. Phase curve mapping is a technique that researchers like

---

[4] https://archive.stsci.edu/missions-and-data/kepler
[5] The stability and mass transfer between two closely orbiting objects are determined by the gravitational forces in binary systems, which are described by the Roche lobe potential.

[20] have used to investigate close-in large planets, also known as "Hot Jupiters". These investigations uncover the planets' thermal and reflective spatial patterns, and phase changes aid in the derivation of specifics such as the longitudinal offsets of the hotter emission zones. The importance of this mapping component lies in its ability to shed light on atmospheric dynamics and the rotational synchronization of these planets with their orbits. The postulated synchronized rotation of hot planets with their orbits, which establishes a direct correlation between their orbital phases and measured longitudes, makes them especially fascinating. While this assumption greatly facilitates mapping efforts, it also creates difficulties in precisely calculating these planets' rotation rates. Researchers such as [21] have highlighted that the influence of atmospheric winds renders traditional approaches like Doppler broadening—which are employed for younger, larger planets on wider orbits—less effective for Hot Jupiters. The influence of stellar tides, which gradually align the rotation periods of these planets with their orbital periods, is projected to cause the synchronization of planetary rotations with their orbits, especially for Hot Jupiters. Mature planetary systems usually match this synchronization, giving mapping investigations a predictable framework. However, to further investigate the possibilities of upcoming observations, mapping techniques are always being improved. The wider implications and potential of planetary mapping are discussed by [22], who highlight the novel insights into the composition and behavior of exoplanets that may be gained from the distinctive brightness structures these techniques reveal. According to these researches, it's critical to recognize the drawbacks and possibilities of eclipse mapping in order to improve our knowledge of planetary atmospheres and architectures. Thus [23]–[27], by providing details about intricate atmospheric and rotational properties that would be difficult to detect directly, planetary mapping advances both our knowledge of exoplanets and the larger discipline of astronomy.

## 4. Materials and Methods

The fundamental objective of the eclipse mapping design is to provide users with the versatility to automate different portions of the code. This allows for the testing of data with multiple parameters and the execution of intricate analytical procedures using straightforward code instructions. For researchers wishing to use computer analysis to investigate various traits and behaviors inside binary star systems—this flexibility is crucial. The program is organized using object-oriented programming techniques and is composed of two primary classes—the Binary class and the Processes class. A key component in gathering and preserving important information about any binary system is the Binary class. Among the details it contains is the orbital model, which is essential to building the eclipse map at every phase point. Furthermore, this class manages the Default map and weighting coefficients. With this configuration, maintaining and referencing core data—which is essential for precise binary system modelling and analysis—can be done more efficiently. Alternatively, the Processes class is meant to contain standard analytical techniques that may be used with the datasets that are kept in the Binary class. The Processes class's present implementation mostly employs static methods and does not require parameters when using the __init__ method to instantiate objects. This design decision makes the class easier to use, but it might restrict customization options for custom analysis scenarios. If specified meta parameters are added to the Processes class at initialization, future versions of the program may benefit from this. The breadth and depth of studies that might be conducted would be increased by this improvement, which would enable users to more precisely customize the analytical procedures to particular research requirements. A major increase in the program's capacity to manage a wide range of analytical tasks would result from this modification, giving researchers access to a more potent tool. At the moment, only Linux and Mac users can use the eclipse mapping software since it depends on methods from the Phoebe[6] Python package. Researchers who predominantly utilize Windows-based systems will find the program's usability limited by this requirement. The removal of these requirements and the development of a program version that fully supports Windows environments would be advantageous in addressing this issue. For now, this program requires Windows users to install Python on a Linux subsystem in order to operate it. After that, they can integrate Python as a remote interpreter in specific development environments. It would improve program accessibility and user-friendliness to eliminate the necessity for such workarounds. The eclipse mapping program can be made more flexible and useful for the astrophysical community by expanding its support for other operating systems and improving the analytical classes' adaptability. By making these enhancements, the research on binary star systems and their dynamics could become more widely adopted and possibly provide more important findings.

### 4.1 Binary Class

An essential component of a Python program created for binary star system analysis is the Binary class. This class can be initialized more than once to enable the simultaneous study of many systems through the use of looping and parallel processing. It maintains data for a particular binary system. Every Binary object point to a directory that contains necessary files, like *parameters.csv*, which contains system parameters, and *light_curve_data.csv*—which has observed light curve data. One of the crucial variables handled in *parameters.csv* is *radius_secondary_given*, a Boolean value that indicates whether or not the software should calculate the radius of the secondary star. This is set to true by default, despite the fact that the estimating feature is not yet implemented. We assume zero eccentricity, which is consistent with the properties of accreting dwarf novae. The accretion grid's resolution is set to $N^2$ by the parameter N, which specifies the pixel length of one axis. The algorithm's chi-squared restriction for each degree of freedom is guided by the *chi_AIM* value, which is usually 0.6. The orbital period, the system name that output directories are produced under, the radii of the primary and secondary stars, the orbital inclination, and the kiloparsec distance between Earth and the binary are additional characteristics. The observed brightness, the accompanying uncertainties, and the Julian Date of each observation are all listed in columns of the *light_curve_data.csv* file. Utilizing this data, the Binary class models the orbital dynamics and light curve of the binary system. The *build_orbital_model()* method establishes functions that return the primary and secondary orbits' *xyz* coordinates with respect to the point for each phase, and this is where the Binary class's functionality starts. The program's automation and integration are facilitated by the storage of these coordinates as features of the Binary object. The method *construct_phasepoints* works with the light curve data to create a set of phase points for every data point that is observed. These phase points are essential for building the visibility matrix later on. By using this technique, one can update a $phase_0$ value, which usually indicates the secondary minimum, by measuring the phase gap between subsequent data points. The program modifies pixel coordinates to correctly reflect the orbital rotation and inclination in order to account for pixel placements throughout the orbit. As a result of this transformation, the accretion grid—which is located inside the orbital plane—maintains a consistent orientation with respect to the secondary star by rotating in sync with the binary system. The last analytical step is to solve for the accretion disc picture by building the entropy equation. To do this, an entropy function is defined using the brightness grid and the weights obtained via the *get_wjks* method. After determining the most likely condition of the accretion disc, the program minimizes this function using Scipy's minimize method, creating a two-dimensional image that shows the brightness distribution of the disc. With the integration of complex data processing and computational modelling techniques, the Binary class offers an all-encompassing analytical tool for simulating and analyzing the physical and observable features of binary star systems.

---

[6] http://phoebe-project.org/

**4.2 Processes Class**

Simplifying the analysis of binary star systems is one of the primary purposes of the Processes class in the eclipse mapping software. Its main function is to schedule the invocation of methods from the Binary class in a particular order so that users can complete in-depth data analysis with just a few simple lines of code. Data loading, computation, and storage related to binary star systems are significantly streamlined by this effective design. The Processes class performs a number of method calls that are essential for controlling a binary object's lifecycle. As stated in software documentation, the class first creates a binary object by consuming parameters and light curve data. Since it serves as the foundation for all ensuing analysis, this setup is essential. After this first stage, the Processes class runs through all of the required methods. In order to analyze the binary system and produce results that are essential for comprehending the dynamics of the system being studied, this involves intricate computational procedures. Following the execution of these procedures, the output is methodically stored in an organized directory inside the Python file's working environment. For example, outputs are kept under "Outputs/Results", which makes sure that all of the data produced during the analysis are arranged and conveniently available for additional examination or study.

**5. Experimental Analysis**

**5.1 Testing the Solving Methods without Constraints**

Many multivariate equation solution techniques, such as SLSQP and CG, are available through the minimize function of the Scipy library. These techniques are essential for resolving the maximum entropy equation; in astrophysics, for example, they are used to examine the brightness distribution of accretion discs in binary star systems. When addressing problems without restrictions like the chi-squared function, different approaches can produce different outcomes. This variance is important because it could affect how accurate and dependable the simulations are that are used to mimic celestial phenomena. Each strategy must be observed in operation, free from chi-squared limitations, in order to be evaluated for efficacy. If the minimization process starts with a two-dimensional Gaussian distribution, the optimal approach would inherently yield a disk-like solution. A Python loop was designed to go through several resolutions and methodologies in order to systematically examine these approaches. A variety of scenarios were covered and the analysis was improved by using *N* values of 10, 15, and 25 for this testing. Just $N = 15$ findings are shown and discussed, though, in order to keep things short. A variety of techniques showed distinct qualities and efficacy based on the tests that were carried out. In Fig. 1, for example, the algorithm's Gaussian starting point picture was not adequately minimized by the approach. However, it seems that the approach depicted in Fig. 2 holds potential and might be investigated further in a later study. It was ultimately decided to use the SLSQP approach, which is shown in Fig. 3. Multiple considerations informed this decision—of the approaches studied, SLSQP showed the fastest solving time and consistently yielded findings that most nearly matched the disk-like structure predicted in these kinds of analyzes. It is noteworthy to mention that the evaluation was centered on the SLSQP findings' structure instead than the pixel values itself. One intriguing finding from the implementation was that the model's pixel size was connected with the occurrence where the image looked to be inverted at the center. But when a chi-squared constraint was added to the model, the inversion problem went away on its own. These kinds of discoveries are essential to improving our knowledge of the computational methods applied in astrophysics, especially in the modelling and study of intricate systems such as accretion discs. Researchers can increase the precision of their models and, consequently, the interpretations and predictions drawn from them by carefully selecting the appropriate solving techniques and comprehending their subtleties.

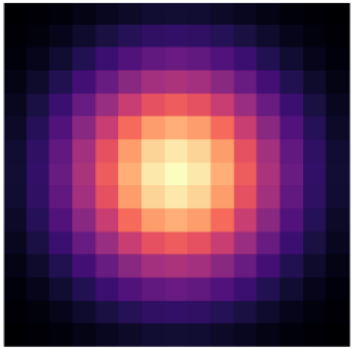

Fig. 1. Unrestricted resolution CG method with $N = 15$

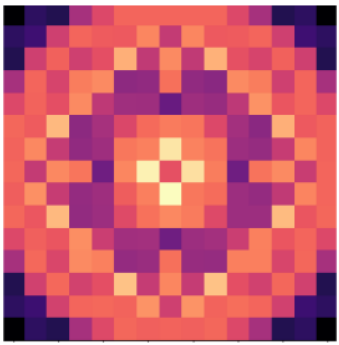

Fig. 2. Unrestricted resolution Truncated Newton method with $N = 15$

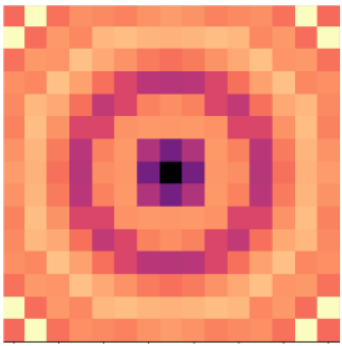

Fig. 3. A solution to the Maximum Entropy Equation that is independent of data, exposing the inherent bias in this approach of solving the equation. Approach: $N = 15$, SLSQP

## 5.2 Light Curve Synthesis and Chi-Square Constraints

The regularized light curve data from the binary star KIC 201325107, whose parameters were previously established using the Phoebe modelling tool as described by [28], were subjected to a thorough examination using the eclipse mapping tool. Using the SLSQP approach for the minimization process was a crucial part of this investigation, as it is an effective way to handle complicated optimization problems. The method was run with different resolution levels ($N = 20$ and $N = 30$) in order to assess how resolution affects the precision and clarity of the resulting images of the accretion disc as shown in Fig. 4. Apart from generating intricate images of the accretion disc, the system further produced a temperature profile for every pixel on the disc. The disk's thermal map, which improves knowledge of its physical properties, was produced by deriving this profile from each pixel's brightness data using the Stefan-Boltzmann formula. To illustrate the brightness of the binary system over time and provide insights into the dynamic processes taking place within the disc, a synthetic light curve graphic was created in conjunction with the temperature profiles. To improve the resolution and image quality of the accretion disc model, the investigation looked into a number of different factors. It was discovered that a value of 3.0 produced the most coherent and smooth disc images. This smoothing factor optimizes the model for a sharper visualization of the disk's structure by controlling the trade-off between detail and noise in the images that are generated. Subsequent experiments involved executing the algorithm at a resolution of $N = 35$, which proved to be the maximum achievable effective resolution within this particular environment. This resolution suggested a departure from the intended disk-like appearance, as the disc image started to show features similar to the diffraction pattern typical of a spherical object. The significance of resolution choices for the accuracy of astrophysical simulations is shown by this observation. By using the algorithm in an anomaly-like epoch, the adaptability of the system was also examined in less perfect circumstances. With this test, we wanted to see how well the algorithm performs when faced with unusual light curve data, as well as how well the maximum entropy equation holds up under difficult conditions. Furthermore, the method was run over a 70-day data set to see how longer observation periods impact the accretion disc image. This greater period aids in the comprehension of the aspects of the disk's temporal evolution and the stability of the accretion processes across time. These thorough experiments and examinations not only show off the eclipse mapping algorithm's potential but also advance our knowledge of accretion disc dynamics and binary star systems in general. Researchers can increase the forecast accuracy and dependability of their models and gain a greater understanding of the intricate dynamics of stellar environments by iteratively improving the algorithm and experimenting with different parameters and situations. The results of this study significantly advance the science of astrophysics, especially when it comes to the study of accretion events and double stars.

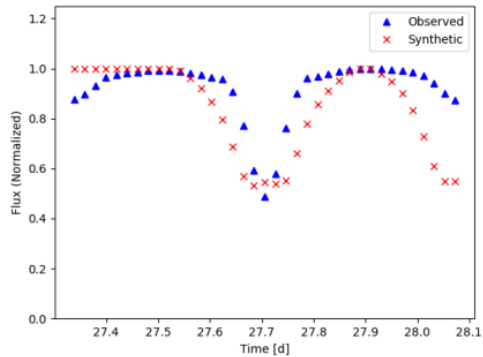

Fig. 4. Observed versus synthetic lightcurve for a single epoch of normalized flux data for KIC 201325107 with SLSQP

**6. Discussion**

There have been some really encouraging results, despite the difficulty in obtaining consistent results—likely caused by code instability that needs more improvement. The technique, for example, successfully identifies a Gaussian hot-spot in the lower left corner of Fig. 5, which illustrates the brightness distribution of an accretion disc and is similar to findings by [16]. This promising conclusion validates the applicability of techniques like the CG algorithm for resolving intricate computational issues in astrophysics, which is encouraging. Even with its accomplishments, the CG approach has had a lot of faults, which could be related to the way the code is designed. However, better results have been obtained when the SLSQP approach is applied with a resolution of $N = 30$. Although deviations in the range of 0 to 0.6 calls for more research, the matching temperature profile displayed in Fig. 6 corresponds quite well with predicted models of accretion discs, indicating a potential association. In an effort to further improve the model, proportionately adding fractions of the flux corresponding to the eclipse region was one way to account for the flux contributions of the secondary light curve. Nevertheless, the artificial light curve did not significantly alter as a result of these modifications, pointing to possible mathematical or programming mistakes. However, when single epoch data is analyzed instead of multi-epoch data, the accretion disc is regularly seen in greater detail. For instance, in Fig. 7, the distributions are better defined, where the Gaussian hot-spot is spread across the right side of the disc. Fig. 8 further illustrates this smearing effect—the averaging effect across all epochs causes the synthetic light curve to mismatch the form of the observed data, resulting in a blurry image. Additional inconsistencies are indicated by the matching temperature profile in Fig. 9, which shows a substantially steeper beginning gradient than the theoretical model. Compared to Fig. 9, the model exhibits less variation even when it deviates from the observed data, as in the case of the anomaly-like solution displayed in Fig. 10. By generating a minima-like solution in Fig. 11, these problems are attempted to be resolved, and the result is a disk-like image with a Gaussian hot-spot, as shown in Fig. 12. On the other hand, it also implies that additional biases can be useful to detect aberrant structures in the disc. Using many layers of analysis [29] – [35] to find phenomena like super-humps in accretion disc images could be one of the future improvements

[36] – [39] in eclipse mapping. Better understanding of the dynamic processes taking place within these celestial structures may be possible with such improvements.

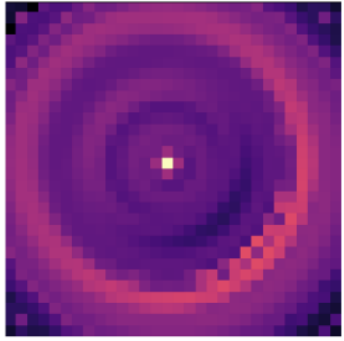

Fig. 5. A maximum entropy solution map corresponding to the synthetic lightcurve with a normalized brightness distribution. For KIC 201325107 with $N = 30$, and Method: SLSQP

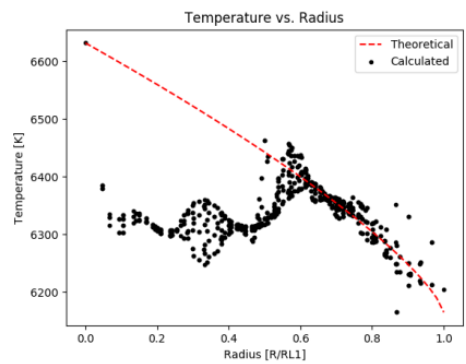

Fig. 6. Temperature profile computed at every pixel and compared to the accretion disk's theoretical temperature profile

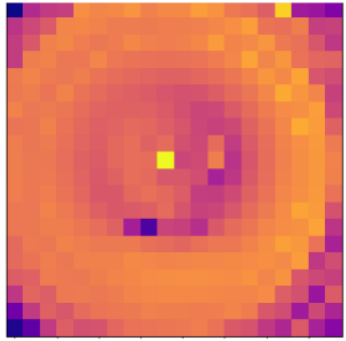

Fig. 7. Smeared accretion image for KIC 201325107 with $N = 20$, and Method: SLSQP

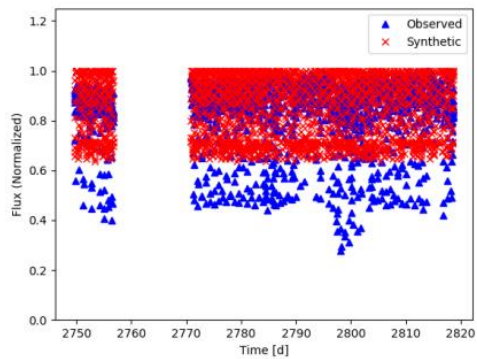

Fig. 8. Observed versus synthetic lightcurve for KIC 201325107

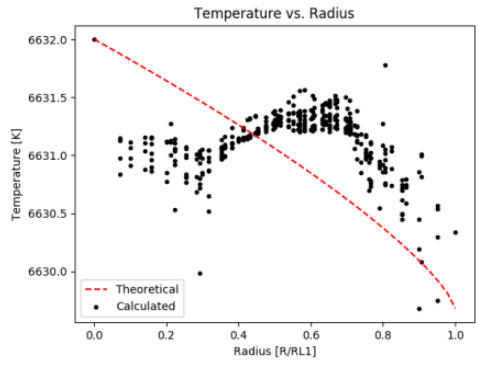

Fig. 9. Smeared temperature profile for KIC 201325107 with $N = 20$, and Method: SLSQP

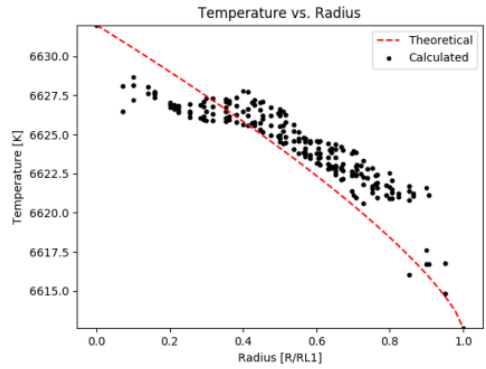

Fig. 10. Temperature profile for KIC 201325107 (anomaly-like) with $N = 20$, and Method: SLSQP

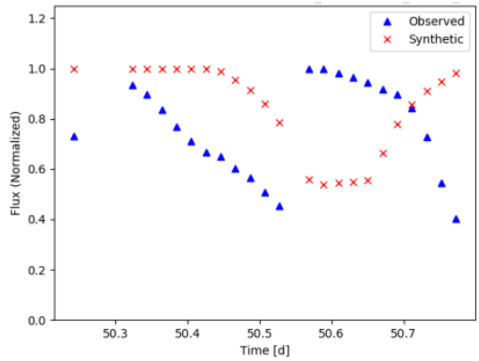

Fig. 11. Observed versus synthetic light curve for KIC 201325107 (anomaly-like) with $N = 20$, and Method: SLSQP

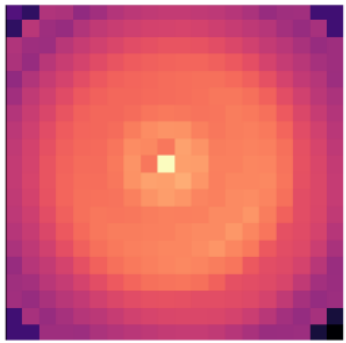

Fig. 12. Accretion image for KIC 201325107 (anomaly-like) with $N = 20$, and Method: SLSQP

## 7. Conclusion and Future Works

The management of biases produced by the weight functions and solving methods [40] – [45] presents a substantial difficulty when employing the maximum entropy approach and eclipse mapping process. For instance, Figs. 5 and 6 illustrate this problem with a clean, oscillatory divergence from the theoretical model in the residuals in the temperature profile from radii 0.1 to 0.55. Fig. 1 illustrates this pattern, which points to a built-in bias in the SLSQP approach. Since it affects the models' accuracy, addressing this bias is essential. We have pointed out that the inherent resonances of SLSQP's unique bias may make it useful for mapping phenomena like super-humps. Using this technique could also make it easier to create three-dimensional models of the thickness of the disc, which would add to the breadth and depth of our knowledge. Moreover, by utilizing the eclipse mapping technique over a wider bandwidth, more thorough maps of accretion discs at multiple wavelength ranges can be produced, offering a larger dataset for examination.

**References**

[1] B. Warner, "High speed astronomy," *Transactions of the Royal Society of South Africa*, vol. 45, no. 1, pp. 1–10, 1983, doi: 10.1080/00359198309520091.

[2] R. V. Shcherbakov, A. Pe'Er, C. S. Reynolds, R. Haas, T. Bode, and P. Laguna, "GRB060218 AS A TIDAL DISRUPTION OF A WHITE DWARF BY AN INTERMEDIATE-MASS BLACK HOLE," *The Astrophysical Journal*, vol. 769, no. 2, p. 85, May 2013, doi: 10.1088/0004-637X/769/2/85.

[3] M. M. Montgomery, I. Voloshina, R. Olenick, K. Meziere, and V. Metlov, "Photometric observations and Numerical modeling of SDSS J162520.29+120308.7, an SU UMa in the CV period gap," *New Astronomy*, vol. 50, pp. 43–51, Jan. 2017, doi: 10.1016/j.newast.2016.07.005.

[4] S. Scaringi, P. J. Groot, and M. Still, "Kepler observations of the eclipsing cataclysmic variable KIS J192748.53+444724.5," *Monthly Notices of the Royal Astronomical Society: Letters*, vol. 435, no. 1, pp. L68–L72, Oct. 2013, doi: 10.1093/MNRASL/SLT099.

[5] N. Kamuni, H. Shah, S. Chintala, N. Kunchakuri, and S. A. O. Dominion, "Enhancing End-to-End Multi-Task Dialogue Systems: A Study on Intrinsic Motivation Reinforcement Learning Algorithms for Improved Training and Adaptability," *2024 IEEE 18th International Conference on Semantic Computing (ICSC)*, pp. 335–340, Feb. 2024, doi: 10.1109/ICSC59802.2024.00063.

[6] N. Kamuni, S. Chintala, N. Kunchakuri, J. S. A. Narasimharaju, and V. Kumar, "Advancing Audio Fingerprinting Accuracy Addressing Background Noise and Distortion Challenges," *2024 IEEE 18th International Conference on Semantic Computing (ICSC)*, pp. 341–345, Feb. 2024, doi: 10.1109/ICSC59802.2024.00064.

[7] N. Kamuni, I. G. A. Cruz, Y. Jaipalreddy, R. Kumar, and V. K. Pandey, "Fuzzy Intrusion Detection Method and Zero-Knowledge Authentication for Internet of Things Networks," *International Journal of Intelligent Systems and Applications in Engineering*, vol. 12, no. 16s, pp. 289–296, Feb. 2024, Accessed: Mar. 31, 2024. [Online]. Available: https://ijisae.org/index.php/IJISAE/article/view/4821

[8] G. S. Kashyap, K. Malik, S. Wazir, and R. Khan, "Using Machine Learning to Quantify the Multimedia Risk Due to Fuzzing," *Multimedia Tools and Applications*, vol. 81, no. 25, pp. 36685–36698, Oct. 2022, doi: 10.1007/s11042-021-11558-9.

[9] M. Kanojia, P. Kamani, G. S. Kashyap, S. Naz, S. Wazir, and A. Chauhan, "Alternative Agriculture Land-Use Transformation Pathways by Partial-Equilibrium Agricultural Sector Model: A Mathematical Approach," Aug. 2023, Accessed: Sep. 16, 2023. [Online]. Available: https://arxiv.org/abs/2308.11632v1

[10] P. Kaur, G. S. Kashyap, A. Kumar, M. T. Nafis, S. Kumar, and V. Shokeen, "From Text to Transformation: A Comprehensive Review of Large Language Models' Versatility," Feb. 2024, Accessed: Mar. 21, 2024. [Online]. Available: https://arxiv.org/abs/2402.16142v1

[11] G. S. Kashyap, A. Siddiqui, R. Siddiqui, K. Malik, S. Wazir, and A. E. I. Brownlee, "Prediction of Suicidal Risk Using Machine Learning Models." Dec. 25, 2021. Accessed: Feb. 04, 2024. [Online]. Available: https://papers.ssrn.com/abstract=4709789

[12] N. Marwah, V. K. Singh, G. S. Kashyap, and S. Wazir, "An analysis of the robustness of UAV agriculture field coverage using multi-agent reinforcement learning," *International Journal of Information Technology (Singapore)*, vol. 15, no. 4, pp. 2317–2327, May 2023, doi: 10.1007/s41870-023-01264-0.

[13] G. S. Kashyap *et al.*, "Revolutionizing Agriculture: A Comprehensive Review of Artificial Intelligence Techniques in Farming,"

Feb. 2024, doi: 10.21203/RS.3.RS-3984385/V1.

[14] S. Wazir, G. S. Kashyap, K. Malik, and A. E. I. Brownlee, "Predicting the Infection Level of COVID-19 Virus Using Normal Distribution-Based Approximation Model and PSO," Springer, Cham, 2023, pp. 75–91. doi: 10.1007/978-3-031-33183-1_5.

[15] G. S. Kashyap, D. Mahajan, O. C. Phukan, A. Kumar, A. E. I. Brownlee, and J. Gao, "From Simulations to Reality: Enhancing Multi-Robot Exploration for Urban Search and Rescue," Nov. 2023, Accessed: Dec. 03, 2023. [Online]. Available: https://arxiv.org/abs/2311.16958v1

[16] R. Baptista, "Eclipse Mapping of Accretion Discs," in *Astrotomography*, Springer, Berlin, Heidelberg, 2007, pp. 307–331. doi: 10.1007/3-540-45339-3_23.

[17] H. A. Knutson *et al.*, "A map of the day-night contrast of the extrasolar planet HD 189733b," *Nature*, vol. 447, no. 7141, pp. 183–186, May 2007, doi: 10.1038/nature05782.

[18] P. K. G. Williams, D. Charbonneau, C. S. Cooper, A. P. Showman, and J. J. Fortney, "Resolving the Surfaces of Extrasolar Planets with Secondary Eclipse Light Curves," *The Astrophysical Journal*, vol. 649, no. 2, pp. 1020–1027, Oct. 2006, doi: 10.1086/506468.

[19] J. De Wit, M. Gillon, B. O. Demory, and S. Seager, "Towards consistent mapping of distant worlds: secondary-eclipse scanning of the exoplanet HD 189733b," *Astronomy & Astrophysics*, vol. 548, p. A128, Dec. 2012, doi: 10.1051/0004-6361/201219060.

[20] K. Menou and E. Rauscher, "ATMOSPHERIC CIRCULATION OF HOT JUPITERS: A SHALLOW THREE-DIMENSIONAL MODEL," *The Astrophysical Journal*, vol. 700, no. 1, p. 887, Jul. 2009, doi: 10.1088/0004-637X/700/1/887.

[21] E. Flowers, M. Brogi, E. Rauscher, E. M.-R. Kempton, and A. Chiavassa, "The High-resolution Transmission Spectrum of HD 189733b Interpreted with Atmospheric Doppler Shifts from Three-dimensional General Circulation Models," *The Astronomical Journal*, vol. 157, no. 5, p. 209, May 2019, doi: 10.3847/1538-3881/AB164C.

[22] N. B. Cowan and E. Agol, "Inverting Phase Functions to Map Exoplanets," *The Astrophysical Journal*, vol. 678, no. 2, pp. L129–L132, May 2008, doi: 10.1086/588553.

[23] H. Habib, G. S. Kashyap, N. Tabassum, and T. Nafis, "Stock Price Prediction Using Artificial Intelligence Based on LSTM–Deep Learning Model," in *Artificial Intelligence & Blockchain in Cyber Physical Systems: Technologies & Applications*, CRC Press, 2023, pp. 93–99. doi: 10.1201/9781003190301-6.

[24] G. S. Kashyap, A. E. I. Brownlee, O. C. Phukan, K. Malik, and S. Wazir, "Roulette-Wheel Selection-Based PSO Algorithm for Solving the Vehicle Routing Problem with Time Windows," Jun. 2023, Accessed: Jul. 04, 2023. [Online]. Available: https://arxiv.org/abs/2306.02308v1

[25] S. Naz and G. S. Kashyap, "Enhancing the predictive capability of a mathematical model for pseudomonas aeruginosa through artificial neural networks," *International Journal of Information Technology 2024*, pp. 1–10, Feb. 2024, doi: 10.1007/S41870-023-01721-W.

[26] S. Wazir, G. S. Kashyap, and P. Saxena, "MLOps: A Review," Aug. 2023, Accessed: Sep. 16, 2023. [Online]. Available: https://arxiv.org/abs/2308.10908v1

[27] G. S. Kashyap *et al.*, "Detection of a facemask in real-time using deep learning methods: Prevention of Covid 19," Jan. 2024, Accessed: Feb. 04, 2024. [Online]. Available: https://arxiv.org/abs/2401.15675v1

[28] "APS -Joint Fall 2017 Meeting of the Texas Section of the APS, Texas Section of the AAPT, and Zone 13 of the Society of Physics Students - Event - Kepler K2 Observations and Modelling of Algol-type binary KIC201325017." https://meetings.aps.org/Meeting/TSF17/Event/311791 (accessed May 03, 2024).

[29] Dodda, S., Kumar, A., Kamuni, N., & Ayyalasomayajula, M. M. T. (2024). Exploring Strategies for Privacy-Preserving Machine Learning in Distributed Environments. Institute of Electrical and Electronics Engineers (IEEE). https://doi.org/10.36227/techrxiv.171340711.17793838/v1

[30] Kumar, A., Dodda, S., Kamuni, N., & Arora, R. K. (2024). Unveiling the Impact of Macroeconomic Policies: A Double Machine Learning Approach to Analyzing Interest Rate Effects on Financial Markets (Version 1). arXiv. https://doi.org/10.48550/ARXIV.2404.07225

[31] Kumar, A., Dodda, S., Kamuni, N., & Vuppalapati, V. S. M. (2024). The Emotional Impact of Game Duration: A Framework for Understanding Player Emotions in Extended Gameplay Sessions (Version 1). arXiv.

https://doi.org/10.48550/ARXIV.2404.00526

[32] Dodda, S., Kumar, A., Kamuni, N., & Ayyalasomayajula, M. M. T. (2024). Exploring Strategies for Privacy-Preserving Machine Learning in Distributed Environments. Institute of Electrical and Electronics Engineers (IEEE). https://doi.org/10.36227/techrxiv.171340711.17793838/v1

[33] Kamuni, N. ., A. Cruz, I. G. ., Jaipalreddy, Y., Kumar, R. ., & Pandey, V. K. . (2024). Fuzzy Intrusion Detection Method and Zero-Knowledge Authentication for Internet of Things Networks. International Journal of Intelligent Systems and Applications in Engineering, 12(16s), 289–296. Retrieved from https://ijisae.org/index.php/IJISAE/article/view/4821

[34] H. Shah and N. Kamuni, "DesignSystemsJS - Building a Design Systems API for aiding standardization and AI integration," 2023 International Conference on Computing, Networking, Telecommunications & Engineering Sciences Applications (CoNTESA), Zagreb, Croatia, 2023, pp. 83-89, doi: 10.1109/CoNTESA61248.2023.10384889

[35] Kamuni, N., Jindal, M., Soni, A., Mallreddy, S. R., & Macha, S. C. (2024). A Novel Audio Representation for Music Genre Identification in MIR (Version 1). arXiv. https://doi.org/10.48550/ARXIV.2404.01058

[36] Chandratreya, Abhijit, et al. "Robotics and Cobotics: A Comprehensive Review of Technological Advancements, Applications, and Collaborative Robotics in Industry." International Journal of Intelligent Systems and Applications in Engineering, vol. 12, no. 21s, 22 Mar. 2024, pp. 1027–1039, ijisae.org/index.php/IJISAE/article/view/5501.

[37] Narne, Suman, et al. "AI-Driven Decision Support Systems in Management: Enhancing Strategic Planning and Execution." International Journal on Recent and Innovation Trends in Computing and Communication, vol. 12, no. 1, 16 Mar. 2024, pp. 268–276, www.ijritcc.org/index.php/ijritcc/article/view/10252. Accessed 6 May 2024.

[38] Dodda, Suresh, et al. "Exploring AI-Driven Innovations in Image Communication Systems for Enhanced Medical Imaging Applications." Journal of Electrical Systems, vol. 20, no. 3s, 4 Apr. 2024, pp. 949–959, journal.esrgroups.org/jes/article/view/1409/, https://doi.org/10.52783/jes.1409. Accessed 30 Apr. 2024.

[39] Dodda, Suresh. "Suresh Dodda: Discussing Automated Payroll Process." Ceoweekly.com, 16 Apr. 2024, ceoweekly.com/discussing-automated-payroll-process-with-suresh-dodda/. Accessed 6 May 2024.

[40] Navin Kamuni, "Enhancing Music Genre Classification through Multi-Algorithm Analysis and User-Friendly Visualization," Journal of Electrical Systems, vol. 20, no. 6s. Science Research Society, pp. 2274–2281, Apr. 29, 2024. doi: 10.52783/jes.3178. Available: http://dx.doi.org/10.52783/jes.3178

[41] J. Thomas, "Optimizing Bio-energy Supply Chain to Achieve Alternative Energy Targets," Journal of Electrical Systems, vol. 20, no. 6s. Science Research Society, pp. 2260–2273, Apr. 29, 2024. doi: 10.52783/jes.3176. Available: http://dx.doi.org/10.52783/jes.3176

[42] J. Thomas, "Optimizing Nurse Scheduling: A Supply Chain Approach for Healthcare Institutions," Journal of Electrical Systems, vol. 20, no. 6s. Science Research Society, pp. 2251–2259, Apr. 29, 2024. doi: 10.52783/jes.3175. Available: http://dx.doi.org/10.52783/jes.3175

[43] Suresh Dodda, "Automated Text Recognition and Segmentation for Historic Map Vectorization: A Mask R-CNN and UNet Approach," Journal of Electrical Systems, vol. 20, no. 7s. Science Research Society, pp. 635–649, May 04, 2024. doi: 10.52783/jes.3413. Available: http://dx.doi.org/10.52783/jes.3413

[44] Arpita Soni, "Advancing Household Robotics: Deep Interactive Reinforcement Learning for Efficient Training and Enhanced Performance," Journal of Electrical Systems, vol. 20, no. 3s. Science Research Society, pp. 1349–1355, Apr. 04, 2024. doi: 10.52783/jes.1510. Available: http://dx.doi.org/10.52783/jes.1510

[45] Hemanth Volikatla, "Reinforcement Learning Approaches for Exploring Discrepancy Variations: Implications for Supply Chain Optimization," Journal of Electrical Systems, vol. 20, no. 5s. Science Research Society, pp. 2861–2868, May 03, 2024. doi: 10.52783/jes.3200. Available: http://dx.doi.org/10.52783/jes.3200